\documentclass[]{aa}  
\usepackage{subfigure}
\usepackage{graphicx}
\usepackage{xcolor}
\usepackage{txfonts}
\usepackage{natbib,twoopt}
\usepackage[breaklinks=true]{hyperref} 
\bibpunct{(}{)}{;}{a}{}{,}             
\makeatletter
  \newcommandtwoopt{\citeads}[3][][]{\href{http://adsabs.harvard.edu/abs/#3}%
    {\def\hyper@linkstart##1##2{}%
     \let\hyper@linkend\@empty\citealp[#1][#2]{#3}}}
  \newcommandtwoopt{\citepads}[3][][]{\href{http://adsabs.harvard.edu/abs/#3}%
    {\def\hyper@linkstart##1##2{}%
     \let\hyper@linkend\@empty\citep[#1][#2]{#3}}}
  \newcommandtwoopt{\citetads}[3][][]{\href{http://adsabs.harvard.edu/abs/#3}%
    {\def\hyper@linkstart##1##2{}%
     \let\hyper@linkend\@empty\citet[#1][#2]{#3}}}
  \newcommandtwoopt{\citeyearads}[3][][]%
    {\href{http://adsabs.harvard.edu/abs/#3}
    {\def\hyper@linkstart##1##2{}%
     \let\hyper@linkend\@empty\citeyear[#1][#2]{#3}}}
\makeatother

\begin{document}

   \title{Average X-ray properties of galaxy groups. \\ From Milky Way-like halos to massive clusters.}


   \author{P. Popesso\inst{1,}\inst{2}\thanks{paola.popesso@eso.org}
   \and
          I. Marini\inst{1}
          \and
          K. Dolag\inst{3,}\inst{4,}\inst{2}
          \and
          G. Lamer\inst{5}
          \and
          B. Csizi\inst{6}
          \and
          V. Biffi\inst{7,}\inst{10}
        \and
        A. Robothan\inst{8}
        \and 
        M. Bravo\inst{9}
          \and
          A. Biviano\inst{7,}\inst{10}
          \and
            S. Vladutescu-Zopp\inst{3}
          \and
          L. Lovisari\inst{11,}\inst{12}
          \and
          S. Ettori\inst{13}
          \and
          M. Angelinelli\inst{13}
          \and
          S. Driver\inst{8}
          \and
          V. Toptun\inst{1}
          \and
          A. Dev\inst{8}
          \and
          D. Mazengo\inst{1}
          \and
          A. Merloni\inst{14}
          \and
          Y. Zhang\inst{14}
          \and
          J. Comparat\inst{14}
          \and
          G. Ponti\inst{11},\inst{14}
          \and
          T. Mroczkowski\inst{1}
          \and
          E. Bulbul\inst{14}
          }

   \institute{European Southern Observatory, Karl Schwarzschildstrasse 2, 85748, Garching bei M\"unchen, Germany \email{paola.popesso@eso.org}
         \and
            Excellence Cluster ORIGINS, Boltzmannstr. 2, D-85748 Garching bei M\"unchen, Germany
        \and
             Universitäts-Sternwarte, Fakultät für Physik, Ludwig-Maximilians-Universität München, Scheinerstr.1, 81679 München, Germany
        \and 
            Max-Planck-Institut für Astrophysik, Karl-Schwarzschildstr. 1, 85741 Garching bei M\"unchen, Germany
        \and
            Leibniz-Institut für Astrophysik Potsdam (AIP), An der Sternwarte 16, 14482 Potsdam, Germany   
        \and
        Universität Innsbruck, Institut für Astro- und Teilchenphysik, Technikerstr. 25/8, 6020 Innsbruck, Austria
        \and
             INAF – Osservatorio Astronomico di Trieste, Via Tiepolo 11, 34143 Trieste, Italy
            \and
            International Centre for Radio Astronomy Research, University of Western Australia, M468, 35 Stirling Highway, Perth, WA 6009, Australia
            \and 
            McMaster University, 1280 Main Street West, Hamilton, Ontario, L8S 4L8, Canada
        \and 
            IFPU – Institute for Fundamental Physics of the Universe, Via Beirut 2, I-34014 Trieste, Italy
        \and
            INAF– Osservatorio Astronomico di Brera, Via E. Bianchi 46, 23807 Merate (LC), Italy
                        \and
            Center for Astrophysics $|$ Harvard $\&$ Smithsonian, 60 Garden Street, Cambridge, MA 02138, USA
        \and
            INAF– Osservatorio Astronomico di Bologna, Via Gobetti 93/3, 40129 Bologna, Italy
                \and
            Max-Planck-Institut für Extraterrestrische Physik (MPE), Giessenbachstr. 1, D-85748 Garching bei München, Germany
             }

   \date{Received September 15, 1996; accepted March 16, 1997}

 
  \abstract
   {In this study, we present the average X-ray properties of massive halos at $z<0.2$ over the largest halo mass range ever probed so far, bridging the gap from Milky Way-like halos to massive clusters.}
   {The results are based on the stacking analysis in the eFEDS area of the GAMA galaxy group sample at $z<0.2$, rigorously tested using a synthetic dataset that mirrors the observed eROSITA X-ray and GAMA optical data based on the lightcones of the Magneticum simulations.}
   {We used a halo mass proxy based on group total luminosity, avoiding systematics linked to velocity dispersion and richness cuts. The stacking is done in bins of halo mass bins and tested in the synthetic dataset for AGN and X-ray binaries contamination, systematics due to the halo mass proxy, and uncertainty in the optical group center. }
   {We provide the average X-ray surface brightness profile in six bins of mass, ranging from Milky Way-like systems to poor clusters at $M_{200} \sim 10^{14}$ $M_{\odot}$. We find that the scatter in the $L_X-M$ relation is driven by gas concentration in groups, as undetected X-ray systems at fixed halo mass exhibit lower central gas concentrations than detected ones, aligning with Magneticum predictions. However, there is a discrepancy regarding dark matter concentration: Magneticum predictions suggest that undetected groups are more concentrated, implying they are older and more relaxed, whereas previous observational findings suggest the opposite. We present new measurements of the $L_{X,500}-M_{500}$ and $L_{X,200}-M_{200}$ relations, from Milky-Way-like halos to massive clusters, filling the gap between these extreme halo mass ranges. Our results indicate that a single power law fits the data across three decades of halo mass, and they align well with previous studies focused on specific halo mass ranges. Magneticum best matches the observed gas distribution across the entire halo mass range, while IllustrisTNG, EAGLE, Simba, and FLAMINGO show larger discrepancies at different mass ranges. This highlights that simulations such as Magneticum, which are not calibrated on $z=0$ galaxy properties, reproduce gas properties well but still lead to overly massive galaxies at the centers of massive halos. Conversely, simulations calibrated on $z=0$ galaxy properties fail to reproduce the gas properties. }
   {This evidence reveals a potential gap in our understanding of the relationship between galaxies and their host structures. Therefore, this work emphasizes the need for a deeper investigation into the connection between gas and dark matter distributions and their impact on central galaxy properties. Such an inquiry is crucial to comprehensively understanding the role and interplay of gravitational forces and feedback-related processes in shaping both the large-scale structure and the galaxy population.}


   \maketitle
%

\section{Introduction}
Galaxy groups are the new frontier of modern cosmology. They represent the crucial test for the predictions of the current models of large-scale structure and galaxy formation and evolution \citep{McCarthy2017}. Their hot gas content, and thus their baryonic mass and X-ray appearance, are predicted to depend on the impact of the feedback of the supermassive black hole (BH) at the center of the central galaxy. At the cluster mass scale, such effects are limited to the central core \citep{LeBrun2014}. In contrast, at the group mass scale, the energy involved is of the same order as the system's binding energy, potentially impacting the entire group volume on a scale of megaparsecs \citep{Oppenheimer2020}.

Modeling the mechanisms—AGN or stellar feedback—that expel baryons from collapsed structures and trigger baryonic exchange represents both the major strength and the greatest weakness of our paradigm of galaxy formation and evolution. Numerical simulations of enormous sophistication and complexity have been dedicated to modeling the interplay between such feedback and gas cooling in massive systems. However, predictions diverge significantly on the amount and distribution of baryonic mass in groups, ranging from exceedingly hot, gas-rich halos to systems completely devoid of gas \citep{Oppenheimer2020}. These discrepancies hinge primarily on the implementation of feedback mechanisms, varying from simpler single-mode scenarios—where a percentage of the supermassive black hole's energy is deposited via a thermal bump into neighboring cells (e.g., Eagle, \citet{Schaye2015}; BahamasXL, \citet{McCarthy2017}; FLAMINGO, \citet{Schaye2023})—to more intricate two-mode implementations. The latter include thermal dumping or black hole-driven outflows at high accretion rates (quasar mode), and kinetic kicks or energy dumping via bubbles at low accretion rates (radio mode) as seen in simulations like IllustrisTNG \citep{pillepich19} and Magneticum \citep{Dolag16}. Navigating through these various solutions and predictions emphasizes the need for stringent observational constraints on the X-ray appearance and hot gas content of galaxy groups.

To date, acquiring such observational constraints has been exceedingly challenging. The Intra-Group Medium (IGrM) within the most common low-mass halos emits primarily through Bremsstrahlung radiation or metal line emissions in the X-rays. Detecting this gas at temperatures below 1-2 keV has posed a formidable challenge for previous X-ray surveys due to limited sensitivity or survey capabilities, particularly below the galaxy cluster mass range \citep{Ponman1996, Mulchaey2000, Osmond2004, Sun2009, Lovisari2015}. To address this issue, researchers have frequently employed stacking techniques using ROSAT All-Sky Survey (RASS) data for galaxy groups and clusters selected through various methods. For instance, \cite{rykoff2008} utilized the {\it{RedMapper}} algorithm on SDSS spectroscopic and photometric data to select clusters based on overdensity and galaxy red sequence, producing an $L_X-M$ relation within $r_{200}$ in the soft X-ray ROSAT band by stacking systems in bins of total optical luminosity or richness. Similarly, \cite{Anderson2015} extended the stacking analysis to Milky Way-like systems by stacking RASS data for a sample of SDSS Bright Red Galaxies, considered to be central galaxies of massive halos. This analysis provided constraints on the $L_X-$gas temperature ($T_X$) relation within $r_{500}$.

More recently, the eROSITA science verification data from eFEDS and the eROSITA All-Sky Survey (eRASS:1) has showcased the instrument's high sensitivity, particularly in the soft X-ray band (0.2-7 keV) below 1-2 keV \citep{Bulbul24}. The first eRASS:1 catalog, encompassing half of the sky, promises to fill the “group desert” in many X-ray scaling relations \citep{2021Univ....7..139L}. However, deeper analyses from eFEDS reveal that eROSITA, $\sim$10 times deeper than eRASS:1, detects only a small fraction of the galaxy group population below halo masses of $10^{14}$ $M_{\odot}$ \citep{Popesso24}. This shortfall suggests that non-gravitational processes, such as AGN feedback, may efficiently expel gas beyond the group’s virial radius, reducing hot gas density and, consequently, the group's X-ray luminosity, since $L_X$ is proportional to the square of gas density.

Stacking analyses of optically selected galaxy groups in eFEDS data reveal lower surface brightness compared to their detected counterparts at fixed halo mass. Synthetic eROSITA data, derived from Magneticum hydrodynamical simulation light cones, indicate that eROSITA's selection function is biased against galaxy groups with lower surface brightness profiles and higher core entropy \citep{xu18,kafer19,xu22,Marini2024a}. This finding supports earlier results by \citet{Popesso24}, although concerns persist regarding potential contamination in optically selected groups and uncertainties in halo mass measurements based on the velocity dispersion of GAMA galaxy groups \citep{Robotham2011} used by \citet{Popesso24}.

Furthermore, \cite{zheng23} and \cite{Zhang24a} conducted stacking analyses on eROSITA data using the optically selected catalogs of \cite{Yang2021} and \cite{tinker21}, respectively, with binning in halo mass proxies provided in the respective catalogs. However, these studies yielded some apparent discrepancies, likely related to systematics in the stacking techniques and the used halo mass proxy.

To better control these systematics, as detailed in \citet{Marini24b}, we generated synthetic optically selected galaxy catalogs from the same light cones used for creating the synthetic eROSITA observations in \citet{Marini2024a}. By mimicking the GAMA selection criteria and employing the group finder of \citet{Robotham2011}, we created a mock galaxy group sample that replicates the GAMA galaxy groups used as a prior catalog for the stacking analysis in \cite{Popesso24}. Leveraging this synthetic dataset, \citet{Popesso2024b} refined the stacking technique, highlighting potential selection effects and systematics. We describe here the results based on this extensive testing to provide robust insights into the average X-ray properties of galaxy groups down to Milky Way-sized halos. In addition, we provide provide the $L_X-M$ scaling relations, within $r_{500}$ and $r_{200}$, bridging the gap between very low mass MW-like groups to massive clusters an so filling a gap in the current literature. We compare our results with previous results in the literature and provide a comprehensive comparison with the predictions of state of the art hydrodynamical simulations such as {\it{Magneticum}} \citep{Dolag16}, IllustrisTNG \citep{pillepich19}, FLAMINGO \citep{Schaye2023}, EAGLE \citep{Schaye2015} and Simba \citep{dave2019}.

The paper is structured as follows: Section 2 describes the observed dataset, including the GAMA optically selected group sample, and details of the eFEDS dataset. Section 3 covers the corresponding synthetic dataset, comprising the galaxy catalog, the GAMA-like galaxy group sample, and the synthetic eROSITA observations. Section 4 discusses the possible selection effects and systematics due to the halo mass proxy, the optical center, and the contamination by AGN and XRB in the stacked average X-ray surface brightness profile. Section 5 presents the results of the X-ray stacking analysis and assesses their robustness. Section 6 analyzes the $L_X-M$ relation within $r_{500}$ and $r_{200}$ and compares it with other works in the literature and predictions of state-of-the-art hydrodynamical simulations. Section 7 provides a summary, discussions, and conclusions of our results.

Throughout the paper, we assume a Flat $\Lambda$CDM cosmology with $H_0 = 67.74$ km~s$^{-1}$~Mpc$^{-1}$ and $\Omega{_m}(z = 0) = 0.3089$ \citep{Planck2016}.

\section{The observed dataset }


\subsection{The GAMA optically selected galaxy group}
The optically selected GAMA group sample \citep{Robotham2011} comprises about 7500 galaxy groups and pairs identified in the spectroscopic sample of the GAMA spectroscopic survey \citep{Driver2022}. This reaches a completeness of $\sim 95\%$ down to the magnitude limit of $r=19.8$. The a Friends-of-Friends (FoF) algorithm described in \citet[][hereafter R11]{Robotham2011} identifies the galaxy groups and pairs. 

Once galaxy membership is identified, the mean group coordinates and redshift are iteratively estimated for each system. The total mass of the systems ($M_{\mathrm{fof}}$) is then estimated from the group’s velocity dispersion ($\sigma_v$) within a variable radius \citep[see][for more details]{Robotham2011}.

However, \citet{Marini24b}, using a mock synthetic catalog based on the same selection algorithm (see next section), show that the halo mass proxy based on velocity dispersion is not a reliable measure for groups with a low number of galaxy members. To address this, a richness cut is required to ensure a minimum number of galaxies to accurately estimate the dispersion, which introduces further selection effects during stacking. \citet{Marini24b} indicate that the total optical luminosity of the groups is the best proxy for the dark matter halo mass. The algorithm effectively retrieves group membership, ensuring that all bright members contributing most to the group’s total luminosity are captured. This accuracy is consistent regardless of group richness, including the case of pairs. Therefore, the halo mass proxy based on total luminosity not only shows the best agreement with the true/input halo mass, but also allows for the creation of a clean group sample without additional selection effects due to richness cuts.

Consequently, we use the provided galaxy membership from the catalog to estimate the total optical luminosity in the r-band following the approach of \cite{Popesso2005}, and then apply the correlation provided there to estimate $M_{200}$. Similarly, we use the best-fit relation from \cite{Popesso2005} to estimate $r_{200}$. Additionally, we derive estimates for $M_{500}$ and $r_{500}$ using the NFW mass distribution model and the concentration-mass relation of \citet{DM14}.

\begin{figure*}
\includegraphics[width=\textwidth]{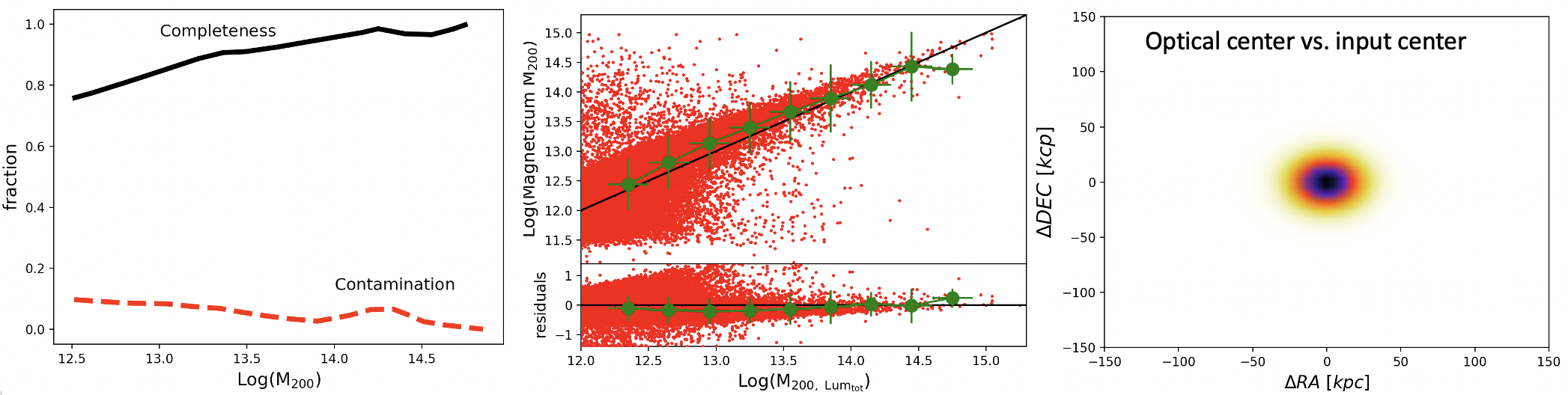}
\caption{{\it{Left panel}}: Completeness and contamination of the optical group catalog based on the Magneticum mock galaxy catalog. {\it{Central panel}}: the top panel shows a comparison between the input halo mass $M_{200}$ and the group mass proxy based on the total luminosity scaled through the scaling relation of \cite{Popesso2005}. The solid line indicates the 1:1 relation. The green symbols indicate the mean $M_{200}$ obtained in bins of 0.3 dex width of $M_{lum}$, whilst the error bar indicates the dispersion. The bottom panel shows the residuals $\Delta=\log(M_{lum})-\log(M_{200})$ of the individual (red) points and the mean values (green symbols). {\it{Right panel}}: the panel shows the difference in $\Delta{RA}$ and $\Delta{Dec}$, estimated in kpc, between the center of the detected optical group and the center of the input halo.}
\label{fig1}
\end{figure*}

\subsection{The eFEDS data}

For this study, we utilized the public Early Data Release (EDR) eROSITA event file of the eFEDS field \citep{Brunner2022}. The field was observed with an unvignetted exposure of approximately 2.5 ks, slightly higher than the anticipated exposure for the future eRASS upon completion, which is about 1.6 ks unvignetted. The dataset contains roughly 11 million events (X-ray photons) detected by eROSITA across the 140 deg$^2$ area of the eFEDS Performance Verification survey. Each photon is assigned an exposure time based on the vignetting-corrected exposure map. Photons in proximity to detected sources from the source catalog are flagged. These sources are classified as point-like or extended according to their X-ray extent \citep{Brunner2022} and further categorized (e.g., galactic, active galactic nuclei, individual galaxies at redshift $z < 0.05$, galaxy groups, and clusters) using multi-wavelength information \citep{Salvato2022, Vulic2022, LiuTeng2022, LiuAng2022, Bulbul2022}.



Despite the smaller volume sampled by eFEDS, the much deeper depth and the very stable background make it still preferable to the much shallower observations of eRASS:1. Thus, we revise the results of \cite{Popesso24} in light of the tests done on mock optically selected catalogs and mock eROSITA observations, to test the contamination and completeness of the prior sample and the reliability of the stacking analysis (see next section). 

\section{The synthetic dataset}

To verify each step of the optical group selection and stacking procedure in the eROSITA data, we construct a mock catalog and a mock eROSITA observation from the same light-cone of the Magneticum simulation, creating an analog of the observational dataset. All details can be found in \citet{Marini2024a} and \citet{Marini24b}. The Magneticum Pathfinder simulation\footnote{\url{http://www.magneticum.org/index.html}} is a comprehensive set of state-of-the-art cosmological hydrodynamical simulations conducted with the P-GADGET3 code \citep[][]{springel_cosmological_2003}. Significant improvements include a higher-order kernel function, time-dependent artificial viscosity, and artificial conduction schemes \citep{dolag_turbulent_2005,beck_improved_2016}. These simulations incorporate various subgrid models to account for unresolved baryonic physics, such as radiative cooling \citep{wiersma_effect_2009}, a uniform time-dependent UV background \citep{haardt_modelling_2001}, star formation and stellar feedback \citep[i.e., galactic winds;][]{springel_cosmological_2003}, and explicit chemical enrichment from stellar evolution \citep{tornatore_chemical_2007}. Additionally, they include models for supermassive black hole (SMBH) growth, accretion, and AGN feedback following established methodologies \citep{springel_cosmological_2003,di_matteo_energy_2005, fabjan_simulating_2010, hirschmann_cosmological_2014}.

Here, we offer a brief description of the synthetic optical and X-ray datasets to emphasize their similarities with the observed data. The description refers to the L30 light cone over an area of 30$\times$30 deg$^2$ up to $z=0.2$, thus simulating only the local Universe analyzed here. Detailed information on the generation of the mock observations is provided in \citet{Marini2024a,Marini24b}.

\subsection{The mock galaxy catalog and galaxy group sample}
\label{optical_mock}
The galaxy mock catalog is derived from the light cones of the Magneticum simulation and it s described in \citet{Marini24b}. The galaxy and halo catalogs within Magneticum are identified using the SubFind halo finder \citep{springel_populating_2001,dolag_substructures_2009}, which compiles a comprehensive list of observables (e.g., stellar mass, halo mass, star formation) by integrating the properties of the constituent particles. The mock galaxy catalog generated from the light-cone is limited to the local Universe up to $z < 0.2$ and covers an area of 30$\times$30 deg$^2$. It includes synthetic magnitudes in the SDSS filters (u, g, r, i, z), observed redshifts, stellar mass, and projected positions on the sky (i.e., RA, Dec) for each galaxy. To simulate a GAMA-like survey, we apply an r-band magnitude cut at 19.8 mag. To mimic observational uncertainties, observed redshifts and stellar masses are assigned errors drawn from Gaussian distributions with $\sigma = 45$ km s$^{-1}$ and 0.2 dex, respectively. Additionally, 5\% of the galaxies are set to undergo catastrophic failure in the spectroscopic survey (i.e., $\Delta v > 500$ km s$^{-1}$), and a spectroscopic completeness of 95\% is simulated.

To create an optically selected galaxy group catalog analogous to the GAMA galaxy group sample, we apply the galaxy group finder algorithm of \cite{Robotham2011} to the mock galaxy catalog after implementing a GAMA-like selection. This approach uses a FoF algorithm for galaxy-galaxy linking, which has been extensively tested on semi-analytic mock catalogs and is designed to be highly robust against the effects of outliers and linking errors. The algorithm's performance has been thoroughly tested in \citet{Marini24b} by matching the input halo catalog with the group catalog in terms of coordinates, redshift, and mass. Here, we report the main results regarding the algorithm's performance in terms of completeness, contamination, and the best halo mass proxy, as these are particularly relevant for the stacking analysis in the corresponding mock eROSITA observations.

As highlighted in \citet{Marini24b}, completeness remains above 90\% down to $M_{200} \sim 10^{13.5} M_{\odot}$, decreases to 80\% at $M_{200} \sim 10^{12.5} M_{\odot}$ (see left panel of Fig. \ref{fig1}). Contamination due to spurious detections remains below 10\% across the whole mass range, also by considering fragmentation \citep[see][for more details]{Marini24b}. These results are qualitatively consistent with those of \cite{Robotham2011}. Quantitatively, the levels of completeness and contamination presented here are higher and lower, respectively, than in \cite{Robotham2011}, due to the lower redshift cut ($z < 0.2$) of the mock galaxy sample compared to GAMA, which extends to $z \sim 0.5$ and the different halo mass proxy used in the analysis.

Indeed, according to \citet{Marini24b}, the best proxy for the input $M_{200}$ is the total optical luminosity ($L_{opt}$). This is estimated by summing up the optical luminosity of the system members down to an absolute magnitude limit, and it is converted into a mass proxy, $M_{lum}$, using a scaling relation. \citet{Marini24b} use the scaling relation of \cite{Popesso2005}. The comparison between $M_{lum}$ and the input $M_{200}$ is shown in the central panel of Fig. \ref{fig1}. While total stellar mass is also a good proxy, it has a slightly larger scatter. The mass derived from velocity dispersion is unreliable for groups with fewer than 10 galaxy members \citep[see][for a more detailed analysis]{Marini24b}.

The scatter of the $M_{lum}-M_{200}$ relation varies with mass, being relatively small ($\sim$0.14 dex) for $M_{lum}>10^{13.5}$ $M_{\odot}$ and increasing up to 0.3 dex at lower values of the proxy with a non-symmetrical behavior (see Fig. \ref{fig1}). This implies that when stacking the optically selected groups in $M_{lum}$ bins, there is contamination by both lower and higher mass groups. \citet{Popesso2024b} study in depth all selection and contamination effects of the scatter of the $M_{lum}-M_{200}$ relation for the \cite{Robotham2011} group catalog. They find that the total luminosity proxy only slightly underestimates the true masses. Nevertheless, this leads to higher contamination by higher mass groups in nearly all halo mass proxy bins \citep[see Table 1 in][]{Popesso2024b}.

The coordinates of the optically selected groups are also in agreement with the input halo coordinated, corresponding to the center of mass. The rms of the residuals shown in the right panel of Fig. \ref{fig1} is about $\sim25$ $kpc$ much smaller than the eROSITA PSF at the considered group redshifts ($z<0.2$, see also next section).

\subsection{The synthetic eROSITA observation}
The photon list of all X-ray emitting components, including hot gas, AGN, and X-ray binaries, is generated using \texttt{PHOX} \citep{biffi_observing_2012, biffi_investigating_2013} for the same light cone L30 used to create the mock galaxy catalog. \texttt{PHOX} computes X-ray spectral emission based on the physical properties of the gas, black hole (BH), and stellar particles in the simulation. The photon list is created for all components up to a redshift $z < 0.2$, providing the X-ray counterpart of the galaxy mock catalog in the local Universe. Detailed modeling of the components can be found in \cite{Marini2024a}.

The synthetic photon lists are used as input files for the Simulation of X-ray Telescopes (\texttt{SIXTE}) software package \citep[v2.7.2;]{dauser_sixte_2019}. \texttt{SIXTE} incorporates all instrumental effects, including the PSF \citep[with $HEW=30$ arcsec, averaged in the 0.2-2.3 keV band as reported in][]{merloni_srgerosita_2024}, redistribution matrix file (RMF), and auxiliary response file (ARF) of the instrument \citep{Predehl2021}. It can also model eROSITA's unvignetted background component due to high-energy particles \citep{LiuAng2022}. We perform mock observations of eRASS:4 in scanning mode using the theoretical attitude file for the three components separately, then combine the event files. In L30, the simulated background in all seven TMs is based on \cite{LiuTeng2022} and represents the spectral emission from all unresolved sources, rescaled to the eRASS:4 depth in line with the simulated emission of the individual X-ray emitting components.

The simulated eROSITA data is processed through eSASS as described in \cite{merloni_srgerosita_2024}. The event files from all emission components and eROSITA TMs are merged and filtered for photon energies within the $0.2-2.3$ keV band. The filtered events are binned into images with a pixel size of $4\arcsec$ and $3240 \times 3240$ pixels. These images correspond to overlapping sky tiles of size $3.6 \times 3.6$ deg$^{2}$, with a unique area of $3.0 \times 3.0$ deg$^{2}$. The L30 is covered by 122 standard eRASS sky tiles. A detailed description of the data reduction is provided in \cite{Marini2024a}, along with the corresponding X-ray catalog of extended and point sources provided by \texttt{eSASS}.

\cite{Marini2024a} thoroughly analyze the completeness and contamination of the extended emission catalog, finding that, after matching the X-ray detections with the input halo catalog, the sample's completeness drops below 80\% at $M_{200} \sim 10^{14} M_{\odot}$, reaches 45\% at $\sim 10^{13.5} M_{\odot}$, and no sources are detected at $\sim 10^{13} M_{\odot}$. The contamination is negligible in the cluster mass range and is about 20\% for $10^{13} M_{\odot} < M_{200} < 10^{14} M_{\odot}$. This is in line with the results obtained by the eROSITA Consortium dedicated works \citep{Biffi18, Clerc2018,Seppi2022}. 

\begin{figure*}
\includegraphics[width=\textwidth]{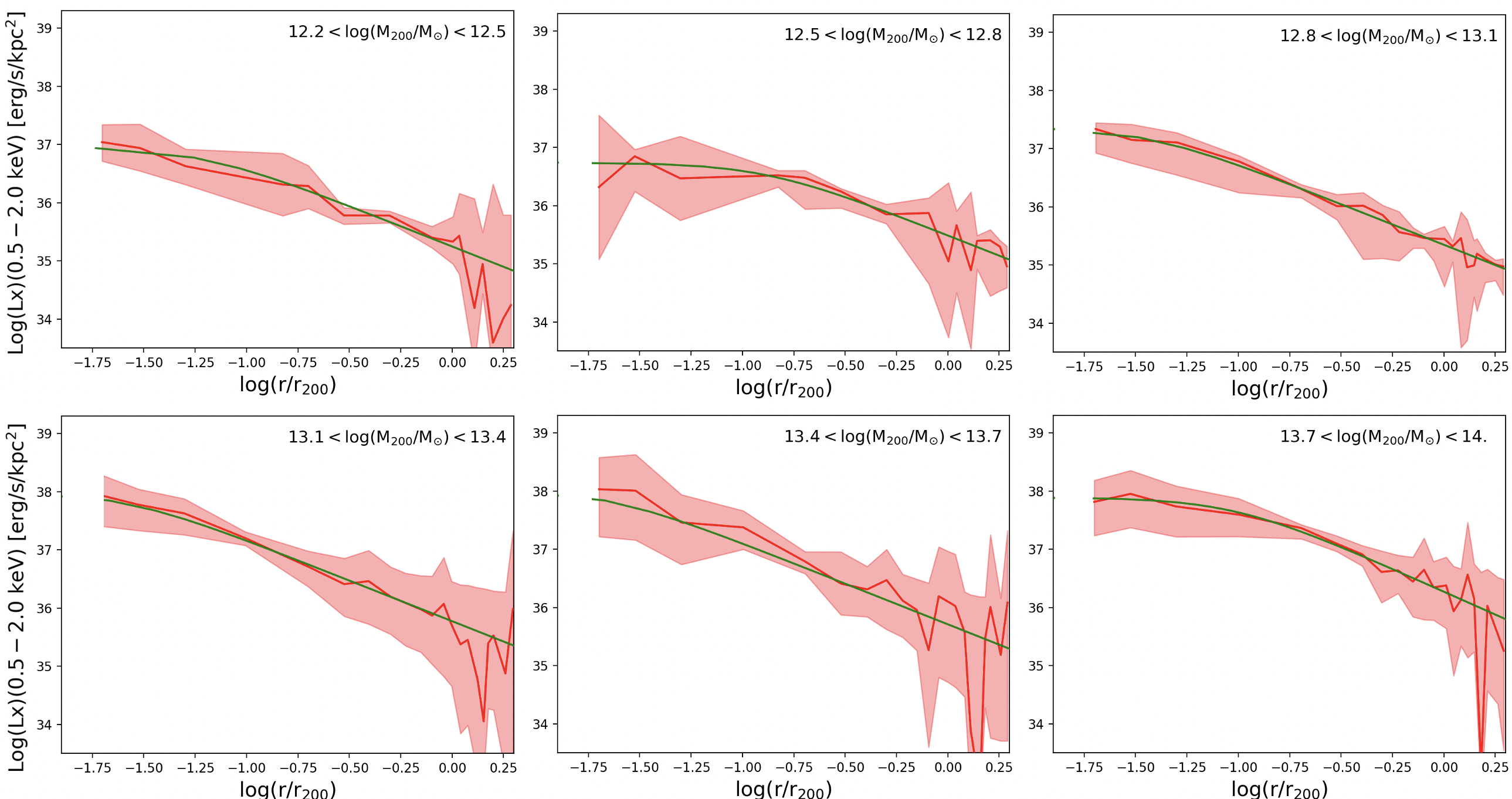}
\caption{The figure shows the X-ray surface brightness profiles in different halo mass bins. The mass bin is indicated in each panel. the mass increases from the left to the right. The red solid line shows the observed profile, while the shaded regions indicate the $1\sigma$ uncertainty estimated via bootstrapping. The green solid line shows the best-fit beta profile with a fixed slope of 0.5.}
\label{average}
\end{figure*}

\section{Testing the stacking procedure}
Before implementing the stacking procedure on the eFEDS data, we validate it using the mock dataset. \citet{Popesso2024b} utilize the mock optically selected group sample, identified via the R11 group finder, as a prior sample for stacking in the mock eROSITA observation of the L30 lightcone. The groups are divided into halo mass bins using the total optical luminosity, the best available proxy for $M_{200}$. Considering the completeness levels (see Section \ref{optical_mock}), the sample is restricted to $M_{200} > 10^{12.2}$ $M_{\odot}$ to target Milky Way-sized halos. The stacking is performed at the optical group coordinates following the procedure outlined in \cite{Popesso24}.

Briefly, the stacking is done by averaging the background subtracted surface brightness profiles within the same annuli around the group center. The background is measured in a region between 2 to 3 Mpc from the group center. All events flagged as point sources in each annulus are excluded and the corresponding annulus area is corrected for the excluded point source area. All groups containing a point source or contaminated by close neighbors within $2\times r_{200}$ are excluded from the prior sample for the stacking. The X-ray luminosity from the stacked signal is derived in the 0.5-2 keV band by selecting only events with a rest frame energy in the selected band at the median redshift of the prior sample. The spectroscopic information is retained and corrected for the effective area to ensure an accurate estimate of the X-ray luminosity. We refer to \cite{Popesso24} for a more detailed description of the procedure.

By applying such a technique for stacking the mock optically selected R11 groups sample on the eROSITA mock observations, \citet{Popesso2024b} identify possible effects of selection biases and uncertainties. We summarize here the results of the analysis:

\begin{enumerate}
\item [-] The completeness and contamination of the prior optical group catalog are such that no significant selection effects are observed (Fig. \ref{fig1} left panel).
\item [-] The most significant systematic effect in the stacking analysis is the scatter between the halo mass proxy and the input halo mass (Fig. \ref{fig1} central panel). Each halo mass bin sample might be differently contaminated by high and low mass systems depending on the level of agreement between the halo mass proxy and the input mass. The net effect is that a bin width of 0.3 dex in $\Delta{M_{lum}}$ corresponds to an actual bin width of 0.45 dex in $\Delta{M_{200}}$ \citep[see][for more details]{Popesso2024b}. This will be considered when plotting the error bars of the mean $M_{lum}$ per bin in further analyses. Because of this, when averaging the Magneticum input PHOX X-ray surface brightness profiles convolved with the eROSITA PSF in bins of $M_{lum}$, this results in overestimated profiles by 0.15 to 0.2 dex for systems with $M_{200}$ below $10^{13}$ $M_{\odot}$. This is mainly due to the lack of a reliable calibration and the large scatter of halo mass proxies in the MW-sized halos in this mass range.
\item [-] The differences in coordinates ($\Delta{RA}$ and $\Delta{Dec}$) between the input halo center and the optical center are much smaller than the eROSITA PSF at the considered redshifts (Fig. \ref{fig1} right panel). Therefore, any mis-centering effect is mitigated by the large FWHM of the eROSITA PSF. We do not observe significant mis-centering relative to the X-ray center in the eSASS extended source detections matched to the input and optically selected group catalogs.
\item [-] When performing the stacking analysis in the mock observations, all these effects tend to compensate for each other to some extent. We confirm an overestimation of the X-ray surface brightness profile for low mass groups below $10^{13}$ $M_{\odot}$ mainly due to contamination in the bins of the halo mass proxy. For halos with masses above $10^{13}$ $M_{\odot}$, we find good agreement between the stacked and input projected PHOX profiles.
\end{enumerate}

\subsection{The AGN contamination}
Correcting for AGN and XRB contamination is essential to properly isolate the IGM contribution. While the contamination of relatively bright AGN ($L_X > 10^{42}$ erg/s in the 0.5-2 keV soft band) can be corrected based on the available eROSITA point source catalog, the contribution of faint and undetected AGN to the X-ray emission of groups can only be corrected by knowing the low-luminosity AGN halo occupation distribution at fixed X-ray luminosity. Indeed, \citet{Popesso2024b} show that, according to the Magneticum simulation, the main contribution to all halos with masses below $10^{13}$ $M_{\odot}$ is due to low-luminosity AGN ($L_X < 10^{41.5}$ erg/s in the 0.5-2 keV soft band). Thus, since these sources are much fainter than the eFEDS detection threshold, the correction must rely on modeling the halo occupation distribution and the conditional X-ray luminosity function of the AGN population, along with the galaxy star formation rate (SFR) distribution.

Many models are available for the halo occupation distribution of AGN at different redshifts \citep[e.g.,][]{Georgakakis2019,krumpe2018,Aird2021,powel2022,krumpe2023,comparat2023}. These models can reproduce the clustering properties and the X-ray luminosity function of the observed AGN population fairly well. However, all models are constrained either by observations at higher redshifts \citep[e.g.,][]{Georgakakis2019,Aird2021,comparat2023} or by relatively bright local AGN \citep[e.g.,][]{krumpe2018,powel2022}. Thus, they require extrapolation to lower redshift, as in our case ($z < 0.2$), or to lower luminosities. Since there is no easy way to discern which model performs better in this parameter range, we rely on the Magneticum results. Indeed, \cite{hirschmann_cosmological_2014} show that the AGN model implemented in Magneticum can broadly match observed black hole properties of the local Universe, including the observed soft and hard X-ray luminosity functions of AGN and clustering properties \citep[see also][]{Biffi18}. \cite{Marini2024a} also shows that the analysis of the eRASS:4-like mock observations based on the Magneticum light cones reproduces well the soft X-ray AGN luminosity function of \cite{marchesi_mock_2020}. Therefore, we correct the X-ray surface brightness profile obtained through the stacking analysis by subtracting a PSF with a signal equal to the percentage of X-ray luminosity due to AGN contamination per halo mass bin, as indicated in Table 2 of \citet{Popesso2024b}.

According to the Magneticum predictions, AGN contamination dominates the X-ray surface brightness profile emission in all halos with masses below $10^{13}$ $M_{\odot}$. These halos contain the majority of low X-ray luminosity AGN in the simulation, consistent with other models in the literature \citep{Georgakakis2019, krumpe2018,Aird2021, krumpe2023}. Consistent with \cite{Popesso24}, AGN contamination is limited to the activity of the central galaxy and is well modeled with a PSF rescaled to the percentage of the X-ray emission corresponding to the AGN contribution.

To remove the residual XRB contamination, we apply the following approach. The XRB contribution is primarily due to the mean SFR of the central galaxies. Satellite galaxies in the local Universe tend to be quiescent or highly quenched, occupying regions well below the Main Sequence (MS) of star-forming galaxies \citep{popesso2019}. Thus, their SFR activity contributes very little to the XRB emission, becoming negligible when averaged azimuthally to obtain the surface brightness profile \citep[see also][]{Popesso24}. Central galaxies in MW-sized halos, on the other hand, tend to be MS galaxies, as shown in \cite{popesso2019}. Therefore, since in most cases the optical center coincides with the central galaxy location, the XRB contribution due to the star formation activity of the central galaxy can be modeled by a PSF rescaled to reproduce the X-ray luminosity of this contribution.

To estimate this, we use the scaling relation from \cite{lehmer2016}, which links the X-ray luminosity in the 0.5-2 keV band to the galaxy SFR. We use the mean SFR of the central galaxies in each halo mass bin to estimate the corresponding mean X-ray luminosity. We then subtract a PSF rescaled to this luminosity from the stacked profile to isolate the intra-group medium contribution.

\begin{figure*}
\includegraphics[width=\textwidth]{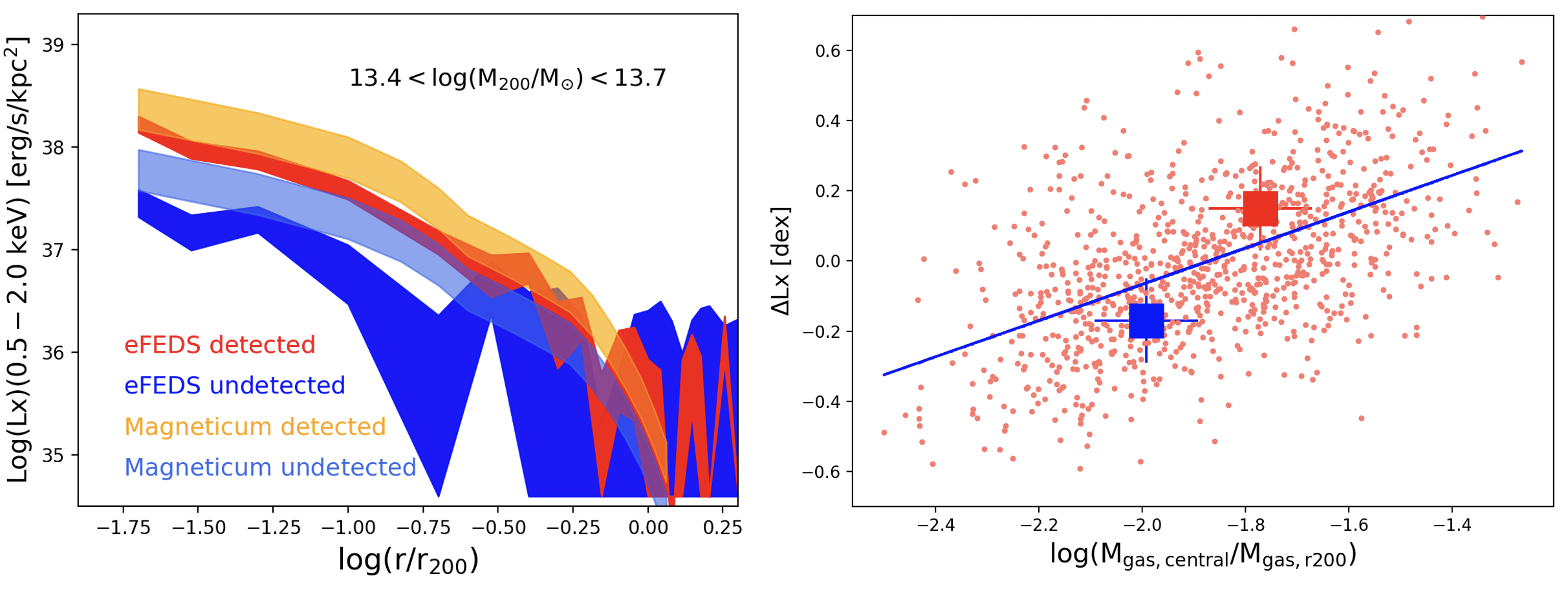}
\caption{{\it{Left panel}}: This panel compares the stacked profiles of eFEDS-detected groups (red-shaded region) and undetected groups (blue-shaded region) within a halo mass bin of $13.4 < \log(M_{200}/M_{\odot}) < 13.7$. Additionally, it displays the stacked profiles from mock eRASS:4 observations for the eSASS-detected groups (orange-shaded region) and the mock R11 undetected groups (light-blue-shaded region) within the same halo mass bin. {\it{Right panel}}: This panel shows the correlation between the residuals ($\Delta{L_X}$) from the \cite{Lovisari2015} $L_{X,500}-M_{500}$ relation and gas mass concentration within the same halo mass bin as in the left panel. The \cite{Lovisari2015} relation has been chosen because it represents a very good fit for Magneticum \citep{Marini2024a,Popesso2024b}. Gas concentration is estimated as the ratio of gas mass within $0.1r_{200}$ to the total mass within $r_{200}$. The red points represent predictions from the Magneticum simulation, while the blue line is the linear best fit. The squares correspond to ratios measured from the observed stacked profiles in the left panel, with the red square indicating the location of detected eFEDS groups and the blue square showing the result for undetected groups.}
\label{det}
\end{figure*}

\begin{figure*}
\includegraphics[width=\textwidth]{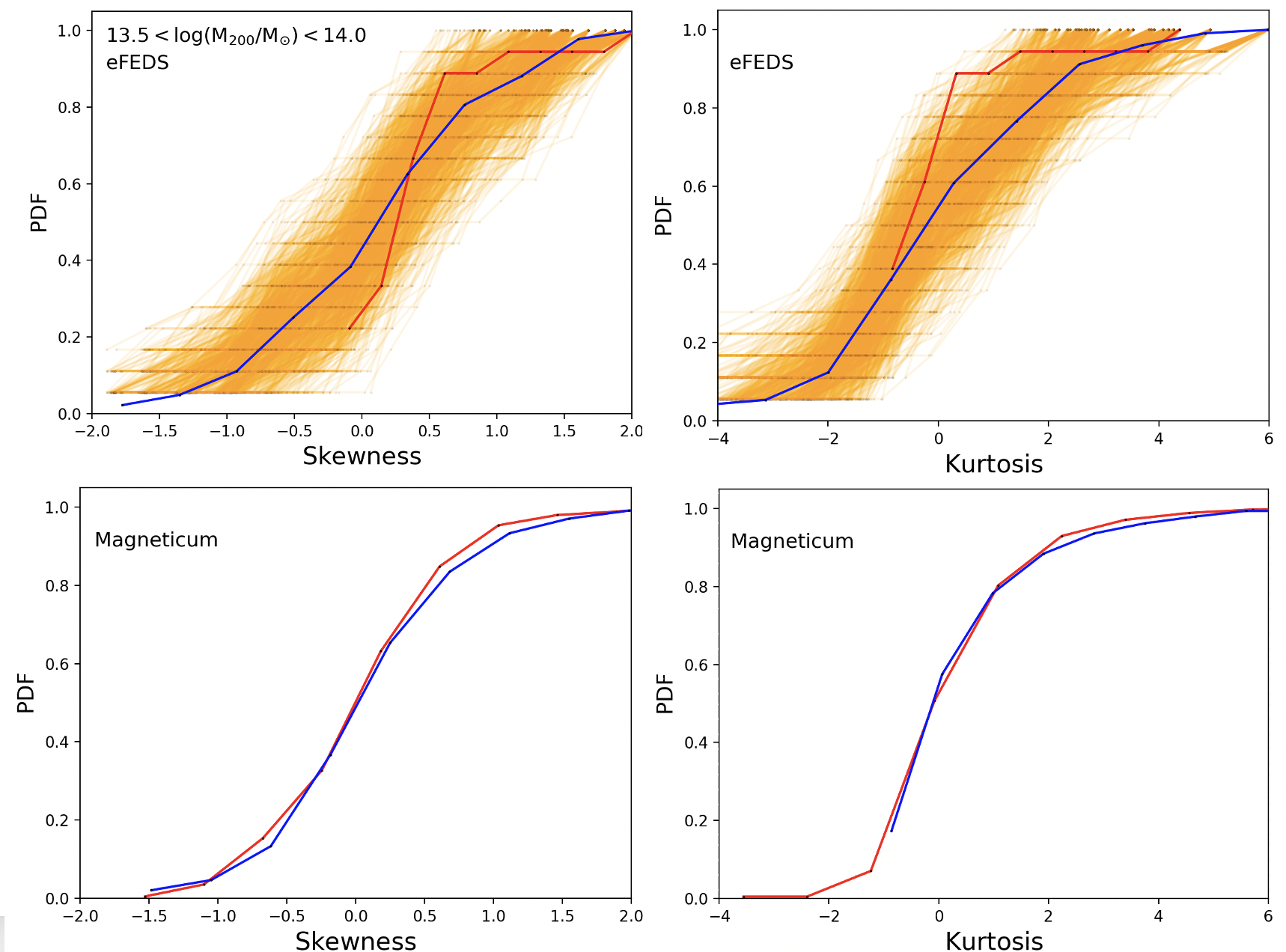}
\caption{{\it{Upper panels}}: The left panel shows the skewness distribution, while the right panel shows the normalized kurtosis distribution (defines as kurtosis$-3$) for detected (red solid line) and undetected (blue solid line) groups in the eFEDS area within the halo mass range $13.5 < \log(M_{200}/M_{\odot}) < 14$. The orange lines represent random extractions from the undetected sample, matching the number of groups in the detected sample within the same halo mass bin. {\it{Lower panels}}: These panels display the skewness (left) and normalised kurtosis (right) distributions for eSASS-detected (red solid line) and R11 undetected (blue solid line) groups within the mock Magneticum dataset, corresponding to the same halo mass bin as in the upper panels.}
\label{kurt}
\end{figure*}

\section{The average intra-group X-ray profiles}
We apply the stacking procedure of \cite{Popesso24}, briefly described above to the eFEDS data. To apply the same AGN correction and consider any selection effects, the stacking is done by using the same halo mass proxy - the total optical luminosity - and the same halo mass binning as used in \citet{Popesso2024b} for the analysis performed on the mock dataset. This, indeed, allows a direct and fair comparison. 

Fig. \ref{average} shows the average X-ray surface brightness profiles obtained in several halo mass bins. We do not distinguish between detected and undetected groups but stack all R11 sources in the mass bins. Only groups with a companion or a point source within $2\times r_{200}$ are discarded from the prior catalog because they would contaminate the background subtraction and the resulting average profile. These account for 6\% of the sample. The profiles are shown in units of $r/r_{200}$. The shaded region in the figure is obtained via bootstrapping in the stacking procedure. All profiles are best fitted by a $\beta$ profile with a fixed slope of 0.5, as indicated in the figure (green solid line). 

\subsection{Detected versus undetected groups}
As in \cite{Popesso24}, we also compare the X-ray surface brightness profiles of detected and undetected groups. This comparison provides valuable insights into the factors driving the scatter in the $L_X-M$ relation, as undetected groups at a fixed halo mass must occupy a region with lower X-ray luminosity than detected ones. Consistent with the findings of \cite{Popesso24}, we identify an optical counterpart for nearly every X-ray detection. Only in a marginal fraction of cases do we observe a mismatch in the group redshift, placing the system outside the redshift window analyzed here \citep[see][for more details]{Popesso24}.

Using a more accurate halo mass proxy, we perform stacking of detected and undetected systems as shown in Fig. \ref{average}. The result for a single halo mass bin ($13.4 < \log(M_{200}/M_{\odot}) < 13.7$) is presented in the left panel of Fig. \ref{det}. We confirm that undetected groups exhibit lower flux and fainter X-ray surface brightness profiles. The difference in mean luminosity at fixed halo mass within $r_{500}$ is 0.43$\pm$0.16 dex, suggesting that the scatter in the relation at this mass must be at least of this magnitude. We also find that undetected systems have a less concentrated average X-ray surface brightness profile compared to detected ones, with less than 33$\pm$15\% of the X-ray luminosity concentrated within the inner 0.2$r_{200}$, compared to 67$\pm$12\% for detected systems. To test the robustness of these findings, we use the available mock dataset.

The mock dataset is analyzed in the same manner as the real observations. The mean profile of the eSASS-detected groups in the mock eROSITA observations within the same halo mass bin is brighter and more concentrated than the average profiles of the undetected groups obtained by stacking the mock R11 galaxy group catalog (see right panel of Fig. \ref{det}). Qualitatively, the Magneticum simulation can replicate the effects observed in the real data. Quantitatively, however, the mock eSASS-detected groups appear brighter than the real ones because they are derived from eRASS:4-like observations, while eFEDS is more than twice deeper. As a result, the distribution between detected and undetected groups differs slightly from that observed due to the brighter detection threshold. Nevertheless, we find good agreement within $1\sigma$. The smaller difference in total luminosity between detected and undetected groups in the mock dataset may also suggest that the scatter predicted by Magneticum at this mass range is smaller than the actual scatter in the $L_X-M$ relation.

When examining the scatter in the $L_X-M$ relation within Magneticum, the simulation indicates that groups at the higher envelope of the relation exhibit a more concentrated gas distribution compared to those in the lower envelope within a fixed halo mass bin. The right panel of Fig. \ref{det} shows the relationship between the residuals in dex from the $L_X-M$ relation of \cite{Lovisari2015}, which aligns with Magneticum predictions, and the gas concentration. This concentration is estimated as the ratio of the gas mass within $0.2 \times r_{200}$ to the mass within $R_{200}$. We observe a clear positive correlation, with a Spearman correlation coefficient of 0.53 and a probability of no correlation of $10^{-4}$.

\begin{figure}
\includegraphics[width=0.5\textwidth]{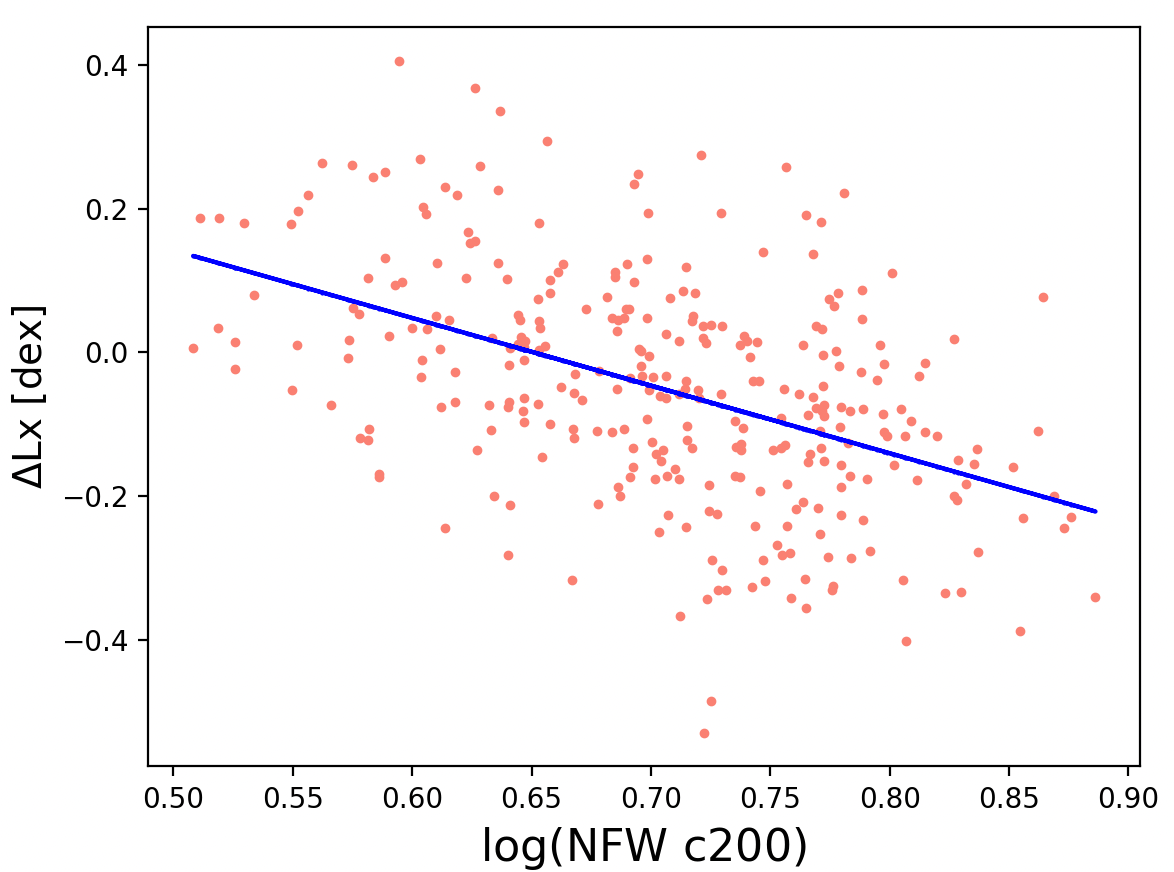}
\caption{Relation between the residuals versus the $L_{X,500}-M_{500}$ relation ($\Delta{L_X}$) of \cite{Lovisari2015} in Magneticum and the concentration of the best fit NFW profile of the dark matter distribution in individual groups. The relation is limited to a halo mass bin at $13.5 < \log(M_{200}/M_{\odot}) < 14$. The solid line indicates the best fit of the anticorrelation.}
\label{con}
\end{figure}

\begin{figure*}
\includegraphics[width=\textwidth]{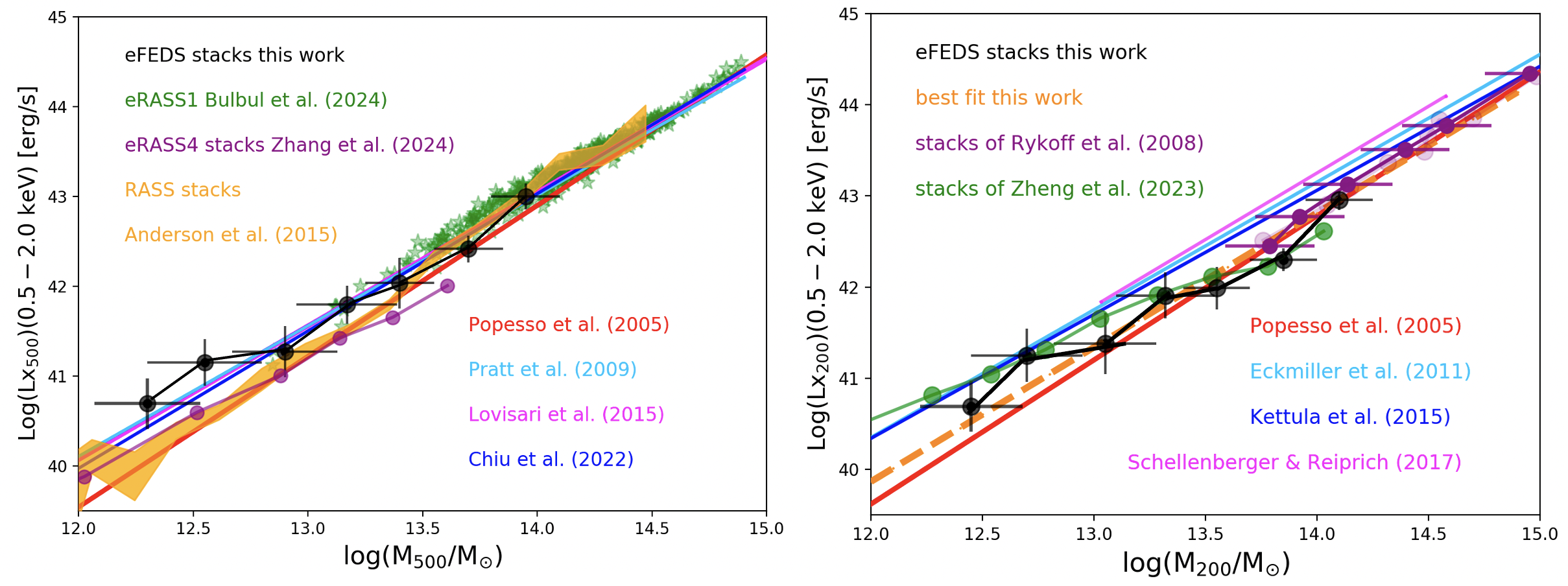}
\caption{{\it{Left Panel}}: The $L_{X,500}-M_{500}$ relation is presented. The black points represent the relation derived by integrating the X-ray surface brightness profiles from Fig. \ref{average} within $r_{500}$. For the highest mass bin, the relation is determined by integrating the average X-ray surface brightness profile from the detections reported by \cite{LiuAng2022}. The green stars correspond to the eRASS:1 cluster sample from \cite{Bulbul24}, while the purple points indicate the stacking results from \cite{Zhang24b}. The orange-shaded region shows the results of the stacking of \cite{Anderson2015}. The red solid line shows the relation from \cite{Popesso2005}, the cyan solid line represents the relation from \cite{pratt09}, the magenta line corresponds to the relation from \cite{Lovisari2015}, and the blue line illustrates the relation from \cite{Chiu2022}. {\it{Right Panel}}: The $L_{X,200}-M_{200}$ relation is displayed. Similar to the left panel, the black points represent the relation obtained by integrating the X-ray surface brightness profiles from Fig. \ref{average}, this time within $r_{200}$. For the highest mass bin, the relation is derived by integrating the average X-ray surface brightness profile from the detections in \cite{LiuAng2022}. The purple points indicate the stacking results from \cite{Rykoff08}, where the filled points refer to binning by total luminosity and the empty purple points refer to binning by richness. The green points show the results from the stacking analysis of \cite{zheng23}. The orange dotted line indicates the best fit of this work at $M_{200}$ larger than $10^{13}$ $M_{\odot}$. The red solid line represents the relation from \cite{Popesso2005}, the cyan line shows the relation from \cite{Eckmiller11}, the magenta line corresponds to \cite{Schellenberger17}, and the blue line represents the relation from \cite{Kettula15}.}
\label{lx_m}
\end{figure*}

We performed the same estimate on the observed X-ray surface brightness profiles of the detected and undetected groups in eFEDS. To achieve this, we estimate the gas mass profile as described in \cite{LiuAng2022}. Specifically, we use a \citet{Vikhlinin09} electron number density model to estimate and integrate along the line of sight the $n_e(r)^2\Lambda(kT, Z)$ profile, where $n_e(r)$ is the electron density profile and $\Lambda(kT, Z)$ is the cooling function depending on the gas temperature and metallicity. The cooling function is derived by assuming a gas temperature from the $M-T_X$ relation of \cite{Lovisari2015} at the mean $M_{500}$ of the halo mass bin and a fixed metallicity of $0.3Z_{\odot}$. The error due to these assumptions is estimated in \citet{Popesso2024b} based on the Magneticum mock observations.

We use the derived gas density profiles to estimate the gas concentration as done for the individual systems in the Magneticum simulation in the same halo mass bin. The result is shown in the right panel of Fig. \ref{det}. The eFEDS detected groups, as shown in \cite{Popesso24}, lie on or slightly above the $L_X-M$ relation in the 0.5-2 keV band and exhibit a higher gas mass concentration than the undetected objects. The location of the points is consistent with the trend predicted by the Magneticum simulation.

Thus, by using a more accurate proxy of the halo mass and a cleaner selection, we confirm the results of \cite{Popesso24} that undetected groups in the eFEDS field exhibit a lower central gas concentration than their X-ray detected counterparts at the same halo mass.

\subsection{Unvirialized systems or AGN feedback effect?}
The lower central concentration of gas in the undetected groups at fixed halo mass compared to the eFEDS detected ones could be explained either by the groups' dynamical state or by non-gravitational processes such as AGN feedback.

The analysis conducted in \citet{Popesso2024b} leads to the creation of a much cleaner galaxy group prior catalog, with a better characterization of systematic effects and a more accurate halo mass proxy than the catalog used in \cite{Popesso24}. Thus, we perform here the same analysis as \cite{Popesso24} on the dynamical properties of the groups to look for higher significance differences between detected and undetected groups. In particular, we check the cumulative distribution of skewness and kurtosis of the individual velocity distributions in a narrow halo mass range ($13.5 < \log(M_{200}/M_{\odot}) < 14.0$) that offers a statistical sample of detected and undetected objects.

The metrics of kurtosis and skewness are two important statistical parameters. The former measures the "tailedness", while the latter the distortion and asymmetry of a distribution. They are, thus, used to qualitatively describe the shape of a velocity distribution, which for a relaxed and virialized system would be consistent with a normal distribution.

A normal distribution has a kurtosis value equal to 3. A distribution with higher kurtosis will have a more tapered (leptokurtik) shape, with a concentration of values around its mean. On the other hand, a distribution with lower kurtosis will have a flatter shape (platykurtik), with values spread over a larger range. The former case indicates the accretion of infalling galaxies creating long tails at high velocities. The latter might indicate a broader distribution in merging systems or projection effects.

The skewness parameter, instead, is more indicative of distortion and asymmetry. A distribution with positive asymmetry will have a longer ‘tail’ to the right of the mean, while a distribution with negative asymmetry will have a longer ‘tail’ to the left of the mean. A perfectly symmetric distribution, such as a normal distribution, would have a skewness equal to zero.

The upper panel of Fig. \ref{kurt} shows the cumulative distribution of the skewness and the normalized kurtosis (kurtosis$-3$, to make it symmetric to the value 0 than the value 3) measured by the R11 algorithm for all groups with more than four members. Since the number of detected groups is significantly smaller than the undetected ones, we extract a random sample from the undetected groups equal to the number of detected systems to compare the consistency of the distributions. We measure the probability that a distribution similar to the detected systems is extracted. This occurs 5\% of the time for skewness and 3\% for kurtosis, leading to a consistency within 2$\sigma$ in both cases. Nevertheless, it is worth mentioning that more than 60\% of the detected systems exhibit a positive skewness, indicating asymmetry, and the majority has a normalized kurtosis lower than 0, indicating broader distributions. This might suggest that the detected systems are undergoing a merging episode or are not fully relaxed. Thus, we conclude that no statistical difference can be observed in the cumulative distributions of the two quantities between detected and undetected systems. However, evidence of unrelaxation or asymmetry is found only in the X-ray-detected systems, while the undetected sample exhibits symmetric skweness and normalized kurtosis distribution.

We repeat the same analysis in the mock R11 galaxy group catalog, where the statistics are much higher for both detected and undetected systems due to the larger area ($\sim$ 900 deg$^2$). As shown in the lower panels of Fig. \ref{kurt}, there is no statistical difference between the distributions of the two quantities in the two samples. This indicates that the dynamical state of the systems in the two samples is statistically similar and does not play a role in the different X-ray surface brightness distributions.

In a first dynamical analysis of the stacked distribution of group galaxy members in the phase-space (velocity dispersion versus group-centric distance), \cite{Popesso24} argue that the lower gas concentration might be due to lower dark matter concentration. Indeed, the analysis conducted with MAMPOSSt \citep{Mamon2013} to solve the Jeans equation for dynamical equilibrium indicates that the undetected groups exhibit a lower concentration than the detected ones. However, if we look at the relation between the residuals from the mean $L_{X,500}-M_{500}$ relation in Magneticum versus the system dark matter concentration obtained by fitting the dark matter particles distribution with an NFW profile, we observe a clear anticorrelation. Fig. \ref{con} shows the relation in the same halo mass bin as in Fig. \ref{det} ($13.5 < \log(M_{200}/M_{\odot}) < 14$). The higher the concentration, the lower the luminosity of the group at fixed halo mass. This indicates that statistically the groups that exhibit a lower gas concentration and remain undetected also show a higher dark matter concentration within $r_{200}$. A higher concentration is indicative of an earlier formation epoch, virialization, and relaxation. This suggests that in Magneticum the undetected groups are more likely relaxed and virialized than the detected ones. Thus, according to the Magneticum predictions, the lower central concentration of hot gas within $r_{200}$ in the lower luminosity and undetected groups cannot be due to a lower dark matter concentration and a different dynamical state but must be due to an alternative non-gravitational process, most likely AGN feedback. We will investigate this aspect in a dedicated paper.

The predictions of Magneticum are at odds with the dark matter concentration obtained with MAMPOSSt in \cite{Popesso24}. However, the use of the halo mass proxy based on velocity dispersion might have contaminated the analysis. A new dynamical analysis of a cleaner and statistically significant group will be the subject of a dedicated paper.

\section{The $L_X$-Mass relation down to MW sized halos}
We estimate the $L_X$-Mass relation down to Milky Way-sized halos within $r_{500}$ and $r_{200}$ by integrating the X-ray surface brightness profiles shown in Fig. \ref{average}. In addition to the six halo mass bins, we also include the average profiles of the detected clusters with $M_{200} > 10^{14}$ $M_{\odot}$ at $z < 0.2$ from eFEDS, using profiles available in the catalog of \cite{LiuAng2022}. As demonstrated in \cite{Popesso24}, the stacked profiles of the detections are consistent with the average profile obtained by averaging the profiles from \cite{LiuAng2022}. This ensures consistency across the entire range of halo masses examined here.

In \citet{Popesso2024b}, we test with the synthetic dataset the mean value and width of the real halo mass distribution corresponding to the halo mass proxy bins used in this work. The results are reported in Table 1 of the same paper. To account for the role of systematics in the use of the halo mass proxy, we adjust the mean halo mass according to the values reported in Table 1 of \citet{Popesso2024b} and assign each halo mass bin an error equal to the bin width reported in the same table for the R11 mock galaxy group sample. This is, indeed, larger than the nominal bin width of 0.3 dex used here due to the underlying distribution of the true Magneticum halo masses. 

The results are shown in Fig. \ref{lx_m} for the two radii, $r_{500}$ (left panel) and $r_{200}$ (right panel). In the left panel, we also include the cluster sample from \cite{Bulbul24} observed in eRASS:1 (green stars). We apply a flux cut at $8\times10^{-13}$ erg sec$^{-1}$ cm$^{-2}$ to ensure a 90\% completeness at $z < 0.1$ for $M_{500} > 10^{14}$ $M_{\odot}$.

The $L_X-M$ relation obtained within $r_{500}$ is consistent within $1\sigma$ over the entire halo mass range with the extrapolation of the observed scaling relations of \cite{pratt09}, \cite{Lovisari2015}, and  \cite{Chiu2022} obtained at higher halo masses. Consequently, we do not provide here an additional best-fit relation. The relation from \citet{Popesso2005}, based on optically selected clusters in the SDSS, is also consistent within $1\sigma$ over the overlapping halo mass range down to $M_{500} \sim 10^{13.5}$ $M_{\odot}$ and deviates by less than $2\sigma$ at lower masses. It is noteworthy that a single power law fits all data points well across nearly three decades of halo masses, from the eRASS:1 massive clusters at $M_{500} \sim 10^{15}$ $M_{\odot}$ to low-mass groups at few times $10^{12}$ $M_{\odot}$. These relations deviate significantly from the self-similar relation, indicating the influence of non-gravitational processes in reducing the amount of hot gas within the considered radius, and consequently, the X-ray luminosity of the groups.

For comparison, we also include the scaling relation from \cite{Anderson2015}. This is derived by stacking massive central galaxies in bins of stellar masses. \cite{Anderson2015} does not provide a measure of $M_{500}$ but of the temperature within $r_{500}$ obtained by converting the $M_{500}$ from the mass-temperature relation of \cite{Sun2009}. We revert the relation to obtain the corresponding $M_{500}$. The stacked relation is in agreement with the relation obtained here within less than $1\sigma$ above $M_{500} > 10^{13}$ $M_{\odot}$ and within $2\sigma$ below this threshold. We also add the $L_X-M$ relation obtained within $r_{500}$ by \cite{Zhang24b}. This relation is consistent with the results presented here within 2.5$\sigma$. The largest discrepancy to ours and \cite{Anderson2015} relation is noted for the most massive groups at $M_{500} > 3\times10^{13}$ $M_{\odot}$. This discrepancy was also noted when comparing the X-ray surface brightness profiles provided by \cite{Zhang24a} and the predictions from Magneticum in \citet{Popesso2024b}. The difference is mainly due to the use of a different halo mass proxy provided by \citet{tinker21}. Indeed, as shown in the Appendix in Fig. \ref{zhang}, when the same halo mass proxy is used, we can reproduce the X-ray surface brightness profile from \cite{Zhang24a} even in the eFEDS area. As discussed in \citet{Popesso2024b}, the halo mass proxy from \cite{tinker21} differs from those explored in this work and, in particular, from the R11 proxy based on the total luminosity. We refer to the Appendix for a more detailed explanation of the discrepancy. 

\begin{figure}
\includegraphics[width=0.5\textwidth]{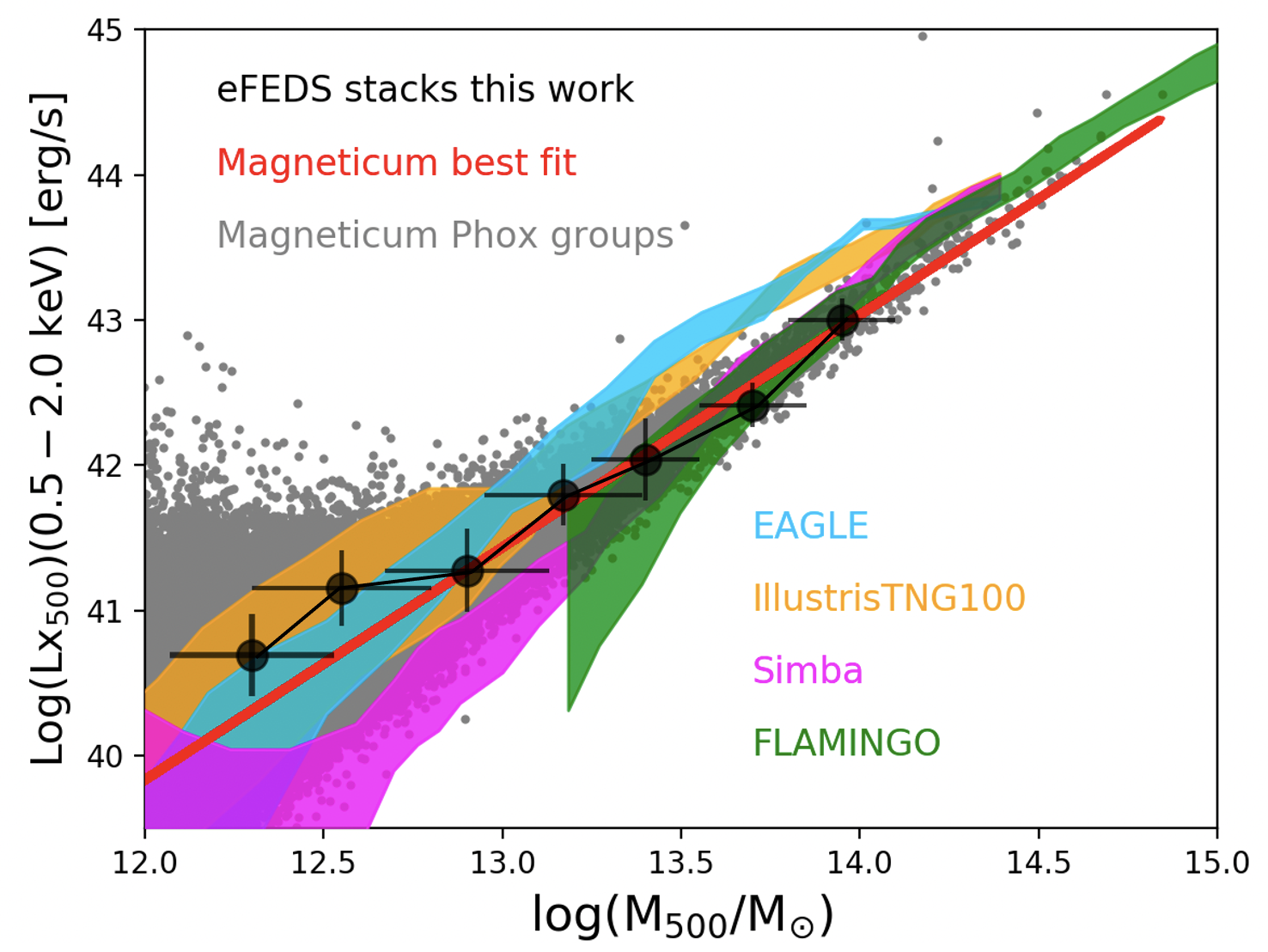}
\caption{Comparison of the $L_{X,500}-M_{500}$ relation presented here with predictions from various state-of-the-art hydrodynamical simulations. The black points represent the relation presented here and shown in the left panel of Fig \ref{lx_m}. The grey points correspond to halos from the Magneticum simulation, where the X-ray luminosity is computed in a manner consistent with the observations by summing the contribution of X-ray photons generated by Phox within $r_{500}$ \citep[see][for details]{Marini2024a}. The red solid line represents the best-fit relation derived from the Magneticum data points.
Additionally, the cyan-shaded region reflects the predictions from the EAGLE simulation \citep{Schaye2015}, while the orange-shaded region shows the predictions from IllustrisTNG100 \cite{pillepich19}. The magenta-shaded region corresponds to the predictions from SIMBA \citep{dave2019}, and the green-shaded region illustrates the results from the FLAMINGO simulation \citep{braspenning2023}.}
\label{sim}
\end{figure}

The right panel of Fig. \ref{lx_m} presents the $L_{X,200}-M_{200}$ relation derived in this study, compared to other works in the literature. Our relation aligns well with stacking results from various optically selected galaxy group samples. The green points depict the stacking analysis by \cite{zheng23}, based on the DESI$+$SDSS optically selected galaxy group sample. This sample utilizes a revised version of the group detection algorithm by \cite{Yang2005}, incorporating both spectroscopic and photometric redshifts \citep[for more details, see][]{Yang2021}. Notably, the results of both stacking analyses show remarkable agreement. This is expected as the \cite{Yang2021} used by \cite{zheng23} provides a halo mass proxy very similar to the one used here and with similar systematics as explored in \citet{Popesso2024b}, which test a mock galaxy group sample based on the \cite{Yang2005} group detection algorithm.  

Additionally, we include results from \cite{rykoff2008}, which are based on the {\it{RedMapper}} cluster selection algorithm applied to SDSS spectroscopic and photometric data at higher halo masses. Our findings are consistent with those of \cite{rykoff2008} within the overlapping mass range. In the figure, filled points represent binning by total optical luminosity, while empty points correspond to binning by cluster richness.

The best-fit result, indicated by the orange dashed line, is based on fitting our data points above a halo mass threshold of $10^{13}$ $M_{\odot}$. This fit follows a power-law form with a slope significantly flatter than the self-similar relation:
\begin{equation}
    L_{X,200}=10^{22.19\pm 0.05} \times (M_{200}/M_{\odot})^{1.47\pm 0.03}
\end{equation}

We do not see evidence for a double power-law over the groups and cluster halo mass range down to $M_{200}=10^{13}$ $M_{\odot}$ as previously suggested in the literature. Indeed, the best fit is perfectly in agreement with the stacking results of \cite{rykoff2008} up to $M_{200}\sim10^{15}$ $M_{\odot}$.

We also compare the $L_{X,500}-M_{500}$ relation obtained in this study with predictions from major hydrodynamical simulations. Fig. \ref{sim} illustrates this comparison. The simulations considered include Magneticum \citep{Dolag16}, EAGLE \citep{Schaye2015}, IllustrisTNG \citep{Pillepich2012}, SIMBA \citep{dave2019}, and FLAMINGO \citep{Schaye2023}. To estimate the X-ray luminosity of the Magneticum groups and clusters, we utilized Phox-generated events, applying the same methodology used for observations within $r_{500}$.

The grey points in the figure represent individual Magneticum groups and clusters, while the red solid line depicts their best-fit relation. This line aligns remarkably well with the relation obtained in this study, being consistent within $1\sigma$ with the scaling relations of \cite{pratt09}, \cite{Lovisari2015}, and \cite{Chiu2022}, as previously demonstrated in \cite{Marini2024a}.

EAGLE, IllustrisTNG, and FLAMINGO simulations generally agree with observations within the group mass range, particularly for Milky Way-like groups. However, they tend to overestimate the X-ray luminosity for more massive groups and poor clusters. In contrast, SIMBA predicts a much steeper relation than observed, overestimating the luminosity of clusters while underestimating it in the low-mass group regime.

In conclusion, the Magneticum simulation best reproduces the $L_{X,500}-M_{500}$ relation and the overall gas properties across the entire halo mass range considered in this study, as also shown in \citet{Popesso2024b}.

\section{Discussion and conclusions}

In this work, we present the latest results of the stacking analysis in the eFEDS area, following extensive testing of the stacking procedure and assessing the selection effects of the prior catalog used. This was achieved by performing each step of the analysis on a synthetic dataset that accurately replicates the observed X-ray and optical data. The main findings of our study are summarized below:

\begin{enumerate}
\item[-] Building on the findings of \citet{Popesso2024b}, we conducted the stacking analysis using a halo mass proxy based on the total luminosity of the group. This approach showed the best agreement with the true mass when tested on the synthetic dataset, allowing us to avoid systematics associated with using the velocity dispersion based on a small number of members and any cuts in richness. The synthetic dataset also enabled us to directly measure contamination from low-luminosity AGN X-ray emission, in line with the Magneticum model of AGN halo occupation and the conditional X-ray luminosity function.
\item[-] We provide X-ray surface brightness profiles of galaxy groups ranging from Milky Way-type systems to the cluster regime ($M_{200}\sim10^{14}$ $M_{\odot}$), with corrections applied for AGN and XRB contamination.
\item[-] Our findings confirm that at a fixed halo mass, systems undetected in eFEDS - likely residing on the lower envelope of the $L_X-M$ relation - exhibit lower central gas concentration compared to detected systems, which populate the upper envelope of the relation. This trend aligns with the predictions of Magneticum for systems at fixed halo mass.
\item[-] Conversely, Magneticum predicts an opposite trend for dark matter concentration: at a fixed halo mass, lower X-ray luminosity corresponds to higher dark matter concentration. This suggests that undetected groups in the simulation are more virialized, relaxed, and older than their detected counterparts, which tend to form later. However, this appears to contradict the findings of \citet{Popesso24}, where a lower dark matter concentration was observed for undetected groups.
\item[-] We provide measurements of the $L_{X,500}-M_{500}$ and $L_{X,200}-M_{200}$ relations, comparing them with other works in the literature based on individual detections and stacked data. We find strong agreement with previous studies and extend the relation down to MW-like systems for both relations. Where discrepancies with previous results arise, we identify the sources of these differences. Our best-fit analysis reveals that a single power law sufficiently fits our data, as well as data points from the literature, providing a good fit over three decades of halo mass. We confirm that across this mass range, the relations at both radii deviate significantly from the self-similar relation, indicating that non-gravitational processes, such as AGN feedback, play a crucial role in shaping the gas distribution and content within halos.
\item[-] Finally, we compare the measured $L_X-M$ relation within $r_{500}$ with predictions from state-of-the-art hydrodynamical simulations. Among the simulations considered in this study, Magneticum emerges as the most accurate in representing the gas distribution and overall content across the entire halo mass range examined here.
\end{enumerate}

While we find a remarkably good agreement between the predictions of Magneticum and the observations, it is puzzling that there is an opposite trend in the dark matter distribution between undetected and detected X-ray systems at fixed halo mass. The difference between these groups offers valuable insight into the underlying scatter of the $L_X-M$ relation and the processes driving it. Our findings suggest that the primary factor influencing this scatter is the gas concentration within halos. However, understanding how this relates to the dark matter distribution is crucial for identifying the mechanisms behind these effects.

If undetected groups exhibit indeed a less concentrated dark matter profile than detected ones, as suggested by \citet{Popesso24}, this would imply that the former are likely younger, still in the process of formation, and less relaxed. This scenario would associate lower gas concentration with dynamical processes such as mergers and gas infall. Conversely, if undetected systems have a more concentrated dark matter profile, as predicted by Magneticum, these groups would likely be older, more relaxed, and shaped predominantly by AGN feedback rather than gravitational processes.

To resolve this discrepancy, it is essential to extend the dynamical analysis conducted in \citet{Popesso24} to a larger statistical sample across the eRASS area. This will be the focus of a series of follow-up papers.

It is also noteworthy that among the simulations considered in this comparison, only Magneticum is not calibrated on the $z=0$ properties and distributions of the galaxy population. Instead, Magneticum is anchored solely to the observed Magorrian relation. While it produces a reasonably accurate galaxy stellar mass function (GSMF), it still faces challenges in stopping star formation in the central galaxies of massive halos. This leads to the creation of overly massive galaxies at the high-mass end of the GSMF \citep{Marini24b}. However, the physics implemented in Magneticum, particularly the modeling of AGN feedback, proves to be highly effective in predicting gas properties across three decades of the halo mass range \citep[see][]{Popesso2024b} in addition to a variety of gas thermodynamical properties within clusters \citep[see also][]{Biffi18,Bahar22}.

In contrast, other hydrodynamic simulations, which are calibrated not only on the Magorrian relation but also on galaxy properties such as the $z=0$ GSMF, successfully reproduce a wide range of galaxy characteristics in the local Universe. However, they struggle to accurately model the distribution and properties of gas within halos. This discrepancy suggests that the physics required to precisely replicate galaxy properties may not simultaneously yield accurate predictions for gas properties, and vice versa. Given that AGN feedback plays a crucial role in these simulations, the inconsistency might stem from differences in how, where, and when energy is released into the medium surrounding galaxies—both at smaller scales (the circumgalactic medium) and at larger scales (the hot halo gas within $r_{200}$ and potentially beyond)—and how gas moves across these scales.

Given that star formation and nuclear activity in galaxies are closely linked to the thermodynamic conditions of the surrounding gas — which serves as the primary fuel for sustaining galactic activity \citep[e.g.][]{2017ApJ...845...80V, Gaspari2017} — this issue may indicate that a fundamental aspect of the connection between galaxies and their host large-scale structures is still missing or misunderstood.
%

\begin{acknowledgements}
PP acknowledges financial support from the European Research Council (ERC) under the European Union’s Horizon Europe research and innovation programme ERC CoG CLEVeR (Grant agreement No. 101045437). AB acknowledges the financial contribution from the INAF mini-grant 1.05.12.04.01 {\it "The dynamics of clusters of galaxies from the projected phase-space distribution of cluster galaxies"}. KD acknowledges support by the COMPLEX project from the European Research Council (ERC) under the European Union’s Horizon 2020 research and innovation program grant agreement ERC-2019-AdG 882679. The calculations for the {\it Magneticum} simulations were carried out at the Leibniz Supercomputer Center (LRZ) under the project pr83li. GP acknowledges financial support from the European Research Council (ERC) under the European Union’s Horizon 2020 research and innovation program Hot- Milk (grant agreement No 865637), support from Bando per il Finanziamento della Ricerca Fondamentale 2022 dell’Istituto Nazionale di Astrofisica (INAF): GO Large program and from the Framework per l’Attrazione e il Rafforzamento delle Eccellenze (FARE) per la ricerca in Italia (R20L5S39T9). SVZ and VB acknowledge support by the \emph{Deut\-sche For\-schungs\-ge\-mein\-schaft, DFG\/} project nr. 415510302

This work is based on data from eROSITA, the soft X-ray instrument aboard SRG, a joint Russian-German science mission supported by the Russian Space Agency (Roskosmos), in the interests of the Russian Academy of Sciences represented by its Space Research Institute (IKI), and the Deutsches Zentrum für Luft- und Raumfahrt (DLR). The SRG spacecraft was built by Lavochkin Association (NPOL) and its subcontractors and is operated by NPOL with support from the Max Planck Institute for Extraterrestrial Physics (MPE).

The development and construction of the eROSITA X-ray instrument was led by MPE, with contributions from the Dr. Karl Remeis Observatory Bamberg \& ECAP (FAU Erlangen-Nuernberg), the University of Hamburg Observatory, the Leibniz Institute for Astrophysics Potsdam (AIP), and the Institute for Astronomy and Astrophysics of the University of Tübingen, with the support of DLR and the Max Planck Society. The Argelander Institute for Astronomy of the University of Bonn and the Ludwig Maximilians Universität Munich also participated in the science preparation for eROSITA.

The eROSITA data shown here were processed using the eSASS software system developed by the German eROSITA consortium.

GAMA is a joint European-Australasian project based around a spectroscopic campaign using the Anglo-Australian Telescope. The GAMA input catalogue is based on data taken from the Sloan Digital Sky Survey and the UKIRT Infrared Deep Sky Survey. Complementary imaging of the GAMA regions is being obtained by a number of independent survey programmes including GALEX MIS, VST KiDS, VISTA VIKING, WISE, Herschel-ATLAS, GMRT and ASKAP providing UV to radio coverage. GAMA is funded by the STFC (UK), the ARC (Australia), the AAO, and the participating institutions. The GAMA website is https://www.gama-survey.org/ .

\end{acknowledgements}

%
%

\bibliographystyle{aa} 
\bibliography{biblio.bib} 

\begin{appendix}

\section{The systematics of the halo mass proxy}

We investigate here the nature of the discrepancy between the stacked data points of \cite{Zhang24b} obtained by stacking the groups identified in the SDSS by \cite{tinker21} in the eRASS:4 data versus our data points and the relations obtained in the literature. 
The $L_{x,500}-M_{500}$ relation of \cite{Zhang24b} is lower compared to the stacked relation obtained here and previously observed scaling relations. This discrepancy was also noted when comparing the X-ray surface brightness profiles provided by \citet{Zhang24a} and the predictions from Magneticum in \citet{Popesso2024b}. The difference is mainly due to the use of a different halo mass proxy provided by \citet{tinker21}. Indeed, as shown in the Appendix in Fig. \ref{zhang}, when the same halo mass proxy is used, we can reproduce the X-ray surface brightness profile from \cite{Zhang24a} even in the eFEDS area. As discussed in \citet{Popesso2024b}, the halo mass proxy from \cite{tinker21} differs from those explored in this work and, in particular, from the R11 proxy based on the total luminosity. 

The left panel of Fig. \ref{tinker} shows the relation between the halo mass proxy used here for the R11 sample and the halo mass proxy provided by \cite{tinker21} for the common sample of SDSS galaxies. The scatter around the one-to-one relation is 0.65 dex, similarly to what foun in \citet{Popesso2024b} in the comparison between the halo masses provided by \citet{Yang2005} and the \citet{tinker21} halo masses. As shown in the right panel of Fig. \ref{tinker}, this implies that a fixed halo mass bin at $13.5 < \log(M_{halo}/M_{\odot}) < 14$ in \cite{tinker21} corresponds to a much broader distribution in R11 masses, with half at masses well below the lower limit of $10^{13.5}$ $M_{\odot}$. We consider that the contamination of lower mass and fainter groups is the main reason for the lower surface brightness profiles obtained in the stacking of \cite{Zhang24a} and for the lower $L_X-M$ relation found in \cite{Zhang24b}.

\begin{figure}
\includegraphics[width=0.5\textwidth]{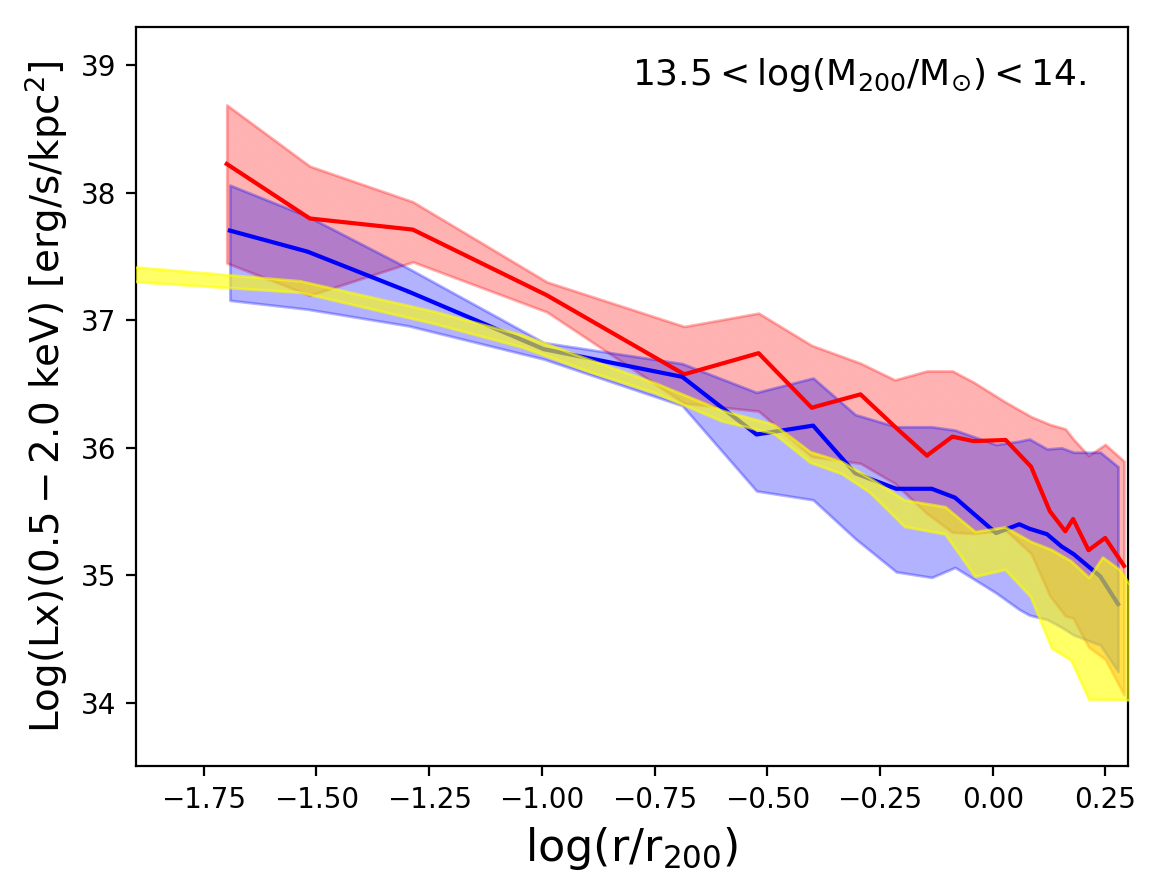}
\caption{The picture shows the comparison between X-ray surface brightness profiles obtained with different prior catalogs. The red solid line shows the average X-ray surface brightness profile obtained from the R11 catalog with the halo mass proxy and procedure used in this paper. The blue solid line indicates the average X-ray surface brightness profile obtained by stacking systems selected according to the halo mass proxy provided by Tinker et al. (2021) and stacked in the eFEDS data. In both profiles, the shaded region indicates the 1$\sigma$ error. The yellow-shaded region shows the profiles obtained by stacking in eRASS:4 the groups selected according to the halo mass proxy of Tinker et al. (2021).}
\label{zhang}
\end{figure}

\begin{figure*}
\includegraphics[width=\textwidth]{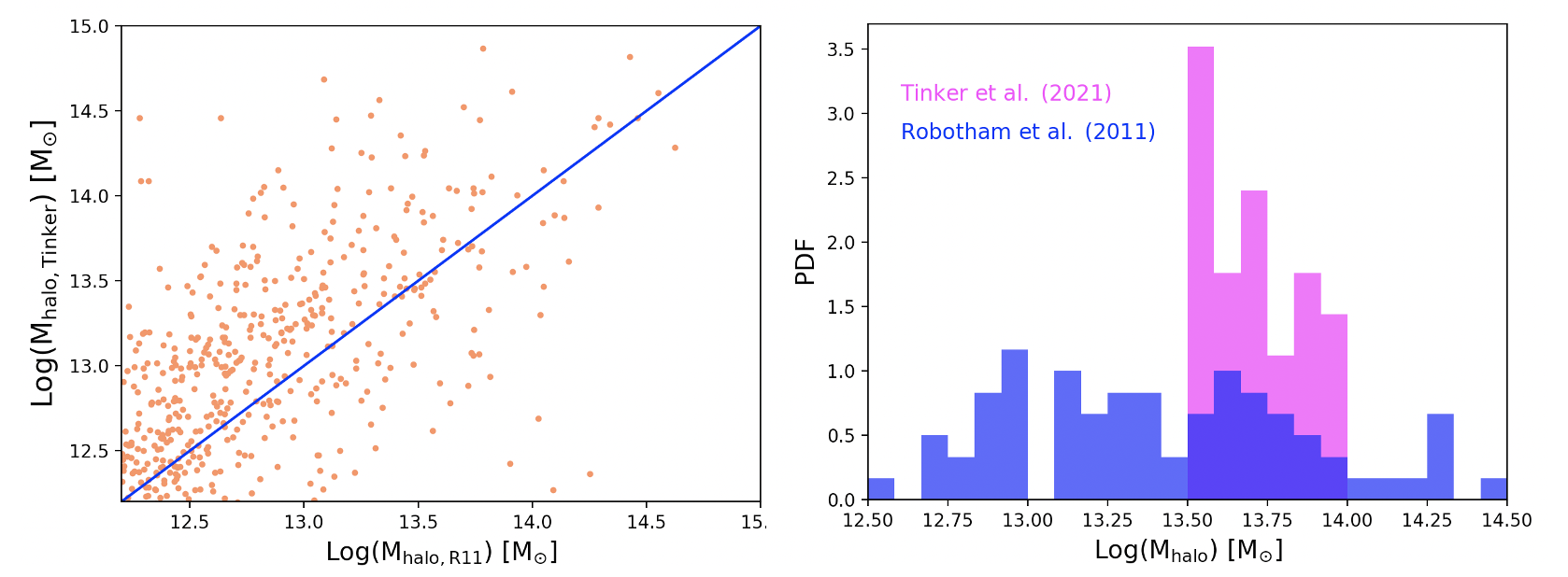}
\caption{Comparison between the properties of the Tinker et al. (2011) group catalog and the R11 catalog. {\it{Left panel:}} Comparison between the halo mass proxy used in the work and based on the total optical luminosity based on the R11 group membership and the halo mass proxy of Tinker et al. (2021). The blue solid line shows the 1:1 relation. {\it{Right panel:}} distribution of the $M_{200}$ halo mass proxy used in this paper (blue histogram) for groups selected according to the halo mass proxy of Tinker et al. (2021) at $13.5 < log(M_{halo}/M_{\odot}) < 14$ (magenta histogram).}
\label{tinker}
\end{figure*}    

\end{appendix}

\end{document}